\documentclass[reprint,amsmath,amssymb,pra,longbibliography]{revtex4-1}

\usepackage{graphicx}
\usepackage{color}
\usepackage{dcolumn}
\usepackage{bm}
\usepackage{mathrsfs}

\newcommand{\Par}{\partial}
\newcommand{\vare}{\varepsilon}

\newcommand{\abs}[1]{\left| #1 \right|}

\newcommand{\cbrc}[1]{\left( #1 \right)}
\newcommand{\brc}[1]{\left[ #1 \right]}
\newcommand{\INF}{\infty}

\newcommand{\RE}{\mathrm{Re}}
\newcommand{\IM}{\mathrm{Im}}

\newcommand{\insMat}[1]{\begin{bmatrix} #1\end{bmatrix}}

\newcommand{\bfm}[1]{\mathbf{#1}}
\newcommand{\Ht}{\widetilde{H}}

\begin{document}


\title{Circularly polarized states and propagating bound states in the
  continuum \\ in a periodic array of cylinders} 

\author{Amgad Abdrabou}
\email{Corresponding author: mabdrabou2-c@my.cityu.edu.hk}
\affiliation{Department of Mathematics, City University of Hong Kong,
  Kowloon, Hong Kong, China}

\author{Ya Yan Lu}
\affiliation{Department of Mathematics, City University of Hong Kong,
   Kowloon, Hong Kong, China}

 \date{\today}

\begin{abstract}
  Bound states in the continuum (BICs) in a periodic structure
  sandwiched between two homogeneous media have interesting
  properties and useful applications in photonics. The topological
  nature of BICs was previously revealed based on a topological charge related to
  the far-field polarization vector of the surrounding resonant
  states. Recently, it
  was established that when a symmetry-protected BIC (with a nonzero
  topological charge) is destroyed by a generic symmetry-breaking
  perturbation, a pair of circularly polarized resonant states (CPSs)
  emerge and the net topological charge is conserved.
    A periodic structure can also support propagating 
  BICs with a nonzero wavevector. These BICs are not protected by
  symmetry in the sense of symmetry mismatch, but they need
  symmetry for their robust existence. Based on a highly accurate 
  computational method for a periodic array of slightly noncircular
  cylinders, we show that a propagating BIC is typically destroyed by a
  structural perturbation that breaks only the in-plane
  inversion symmetry, and when this happens, a pair of CPSs of
  opposite handedness emerge so that the net topological charge is
  conserved. We also study the generation and annihilation of CPSs
  when a structural parameter is varied. It is shown that two CPSs 
  with opposite topological charge and same handedness, connected to
  two BICs or in a continuous branch from one BIC, may collapse and
  become a CPS with a zero charge. Our study clarifies the important
  connection between symmetry and topological charge conservation. 
\end{abstract}
\maketitle

\section{Introduction}
\label{S1}

Bound states in the continuum (BICs) are trapped or guided modes with
their frequencies in the radiation continua~\cite{neumann29,hsu16}.
They exist in a variety of photonic structures including periodic
structures sandwiched  between two homogeneous
media~\cite{bonnet94,padd00,ochiai01,tikh02,shipman03,lee12,port05,mari08,hsu13_2,bulg14b,gan16},
waveguides with local defects~\cite{evans94,bulg08}, waveguides with
lateral  leakage channels~\cite{zou15,bezus18,nguyen19,yu19,byk20},
etc.  
In structures that are invariant or periodic in one or two spatial 
directions, a BIC is a special point with an infinite quality factor 
($Q$ factor) in a band of resonant 
states~\cite{hsu13_2,yuan17_2,yuan18,jin19}, and it becomes a 
high-$Q$ resonance  if the structure  is  perturbed generically \cite{kosh18,yuan20}. 
High-$Q$ resonances lead to strong local field
enhancement~\cite{mocella15,yoon15,huzhen20} and abrupt features in
reflection and transmission spectra~\cite{yuan17}, and are essential
for sensing, lasing, switching and nonlinear optics applications. 

For theoretical interest and practical applications, it is important
to understand how a  BIC is affected by a perturbation of the
structure.
If the BIC is protected by a symmetry \cite{bonnet94,padd00,ochiai01,tikh02,shipman03,lee12,evans94,bulg08}, i.e., there is a symmetry mismatch between the BIC and the compatible radiation modes, it naturally continues its existence if the perturbation preserves the symmetry \cite{bonnet94,evans94,shipman07}. Therefore, a symmetry-protected BIC is robust with respect to symmetry-preserving  perturbations.
If a  perturbation breaks the symmetry, a symmetry-protected BIC
typically, but not always, becomes a resonant state with a finite $Q$
factor~\cite{kosh18,yuan20,yuan20b}. In periodic structures sandwiched
between two homogeneous media, there are also BICs with a nonzero
Bloch wavevector, and they propagate in the periodic directions \cite{port05,mari08,hsu13_2,bulg14b,yuan17,maksim,hu18}.
Such a propagating BIC is not protected  by symmetry in the usual sense, but  can also be robust with respect to symmetry-preserving perturbations~\cite{zhen14,yuan17ol,conrob}.
More precisely, in a periodic structure with an up-down mirror
symmetry and an in-plane inversion symmetry, a generic low-frequency
propagating  BIC  (with only one radiation channel) continues its
existence 
 if the structure is perturbed by a perturbation
preserving these two symmetries~\cite{yuan17ol,conrob}.

The BICs in periodic structures exhibit interesting topological
properties. Zhen {\it el al.}~\cite{zhen14} realized that a BIC  in a
biperiodic structure  is a polarization singularity in momentum
space (the plane of two wavevector components), and defined a
topological charge based on the winding number of the far-field
polarization vector. Since the
far field of a resonant state is typically elliptically polarized,
Bulgakov and Makismov 
refined the definition using the major polarization vector~\cite{maksim}. The topological charge can be used to classify the BICs and illustrate their generation, interaction and annihilation processes when structural  parameters are tuned \cite{zhen14,maksim}. Importantly, the topological charge 
is a conserved quantity that cannot be changed by small structural
perturbations. However, this does not imply that the BICs (with a
nonzero topological charge)  are robust with respect to arbitrary
structural perturbations, because a resonant state can be circularly
polarized and also has a nonzero topological charge. In a recent work,
Liu {\it el al.}~\cite{fudan} showed that symmetry-protected BICs in a
photonic crystal slab, protected by the in-plane  inversion 
symmetry, turn to pairs of circularly polarized resonant states (CPSs)
when the structure is perturbed breaking the symmetry.

Most propagating BICs are found in periodic structures with both the 
up-down mirror symmetry and the in-plane inversion symmetry. 
It is known that if one of these two symmetries is broken, a 
propagating BIC is usually, but not always, destroyed 
\cite{hu18,yuan20b}. It has been shown that CPSs can exist in structures
without the up-down  mirror symmetry and in-plane  inversion
symmetry~\cite{yin20}. In this paper,  we show that CPSs emerge
when propagating BICs are destroyed by generic and arbitrarily small
perturbations that break only the in-plane inversion  symmetry. 
%
%
%
We consider vectorial BICs in a periodic array of of
dielectric cylinders, introduce a small deformation of the cylinder
boundary, and show that pairs of CPSs emerge when propagating BICs
with topological charge $\pm 1$ are destroyed. In addition, we follow
the CPSs as the deformation parameter is further increased, and show
that two CPSs with opposite topological charges and same handedness may merge, and their dependence on the deformation parameter may be multi-valued with self-generation and annihilation points. 


The rest of this paper is organized as follows. In Sec.~\ref{S2}, we
recall the mathematical formulation for vectorial eigenmodes 
in two-dimensional (2D) periodic structures and the definition of
topological charge. In Sec.~\ref{S2a}, we consider a periodic array of
circular cylinders and briefly describe our computational method. 
In Sec.~\ref{S3}, we introduce boundary deformations to the cylinders
and show  the generation and annihilation of CPSs. The paper is
concluded with a brief  discussion in Sec.~\ref{S4}.

\section{Eigenmodes in periodic structures}
\label{S2}

We consider a 2D structure that is invariant in $x$, periodic in $y$ with 
period $L$,  bounded in $z$ by $\abs{z}<D$ for some $D>0$,
and surrounded by air (for $|z| > D$), where
$\{x,y,z \}$ is a Cartesian coordinate system.
Let $\vare = \vare(y,z)$ be the relative permittivity of this
structure and its surrounding medium, then 
$\vare({\bf r}) =\vare(y+L,z)$ for all ${\bf r}=(y,z)$ and 
$\vare({\bf r}) = 1$ for $\abs{z}>D$. We study time-harmonic
electromagnetic waves that depend on time $t$ and variable $x$
as $\exp[ i(\alpha x -\omega  t)]$, where $\omega$ is the angular
frequency and $\alpha$ is a real wavenumber in the $x$
direction. From the frequency-domain Maxwell's equations, it 
is easy to obtain the following system 
\begin{eqnarray}
  \label{Eq1}
&&	\nabla\cdot\cbrc{ \frac{\vare}{\eta}\nabla E_x} +
   \nabla\cdot\cbrc{ \frac{\alpha}{k\eta} P \cdot\nabla
   \widetilde{H}_x}+\vare E_x = 0, \quad \\
  \label{Eq2}
&&	\nabla\cdot\cbrc{ \frac{1}{\eta}\nabla \widetilde{H}_x} -
   \nabla\cdot\cbrc{ \frac{\alpha }{k\eta} P\cdot\nabla E_x}+\widetilde{H}_x = 0,
\end{eqnarray}
where $E_x$ is the $x$ component of the electric field, 
$\widetilde{H}_x$ is the $x$ component of a  scaled magnetic field
(magnetic field multiplied by free space impedance), $k = \omega/c$
is the free space wavenumber, $c$ is the speed of 
light in vacuum, and 
\begin{equation}
  \label{defeta}
\eta = k^2\vare({\bf r}) -\alpha^2, \quad 
   \nabla = \insMat{ \Par_y \\ \Par_z},    \quad  P = \insMat{0& 1\\-1 & 0}.  
\end{equation}
 The other four field components can be obtained from the following equations:
 \begin{eqnarray}
   \label{Eq3}
   && \insMat{E_y\\E_z}=\frac{i}{\eta}\cbrc{ \alpha \nabla E_x + k\,
   P \cdot \nabla  \widetilde{H}_x}, \\
\label{Eq4}
&&	\insMat{\widetilde{H}_y\\\widetilde{H}_z}=\frac{i}{\eta}\cbrc{
   \alpha \nabla \widetilde{H}_x - k\,\vare P \cdot \nabla  E_x}.
 \end{eqnarray}
If $\alpha = 0$, the equations for $E_x$ and $\widetilde{H}_z$ are
decoupled, we say the corresponding wave is scalar. The case $\alpha
\ne 0$ is referred to as vectorial.  
 
Due to the periodicity in $y$, any eigenmode of the structure is a
Bloch mode with an electric field given by 
\begin{equation}
  \label{bloch}
{\bf E}({\bf r}) = {\bf F}({\bf r}) e^ {i\beta y},   
\end{equation}
where $\beta$ is a real Bloch wavenumber satisfying $\abs{\beta}\leq
\pi/L$, and ${\bf F}$ is a periodic in $y$ with period $L$. For
$\abs{z}>D$, the field can be expanded in plane waves as
\begin{equation}
  \label{planewave}
  {\bf E}({\bf r})  =  \sum_{m= -\INF }^{\INF} {\bf c}_m^{\pm} e^{i
    (\beta_m y\pm \gamma_m z)},\quad  \pm z> D, 
\end{equation}
where 
$\beta_0 = \beta$, and
\begin{equation}
  \label{defbeta}
  \beta_m = \beta+ \frac{2\pi m}{L}, \quad \gamma_m =  \sqrt{ k^2 -\alpha^2 
-\beta_m^2}.
\end{equation}
A guided mode satisfies the condition ${\bf E} \to {\bf 0}$ as $z\to \pm 
\INF$. For a positive $k$, guided modes usually
exist when $ k < \sqrt{ \alpha^2 + \beta^2}$, so that all $\gamma_m$ are
pure imaginary and all plane waves in the right hand side of Eq.~(\ref{planewave}) are evanescent
in $z$. A BIC is also a guided mode, but it satisfies the condition
$k > \sqrt{\alpha^2 + \beta^2}$, thus $\gamma_0$ and probably a few
other 
$\gamma_m$ are real positive. We study BICs with a real
$k$, a real $\alpha$, and a real $\beta$ such that 
\begin{equation}
  \label{onerc}
  \sqrt{ \alpha^2 + \beta^2} < k <
  \sqrt{ \alpha^2 + \left( \frac{2\pi}{L} -|\beta| \right)^2}. 
\end{equation}
The above condition implies that $\gamma_0$ is positive and all other
$\gamma_m$ are pure imaginary. Since the 
plane waves $\exp[ i (\beta_0 y \pm \gamma_0 z)]$ can propagate to
infinity, the coefficients ${\bf c}_0^{\pm}$ of the BIC must vanish.

A resonant state is also an eigenmode, but it satisfies an outgoing
radiation condition as $z \to \pm \infty$. Because of the radiation
loss, the amplitude of a  resonant state decays in time, thus the
frequency $\omega$ or $k$ must be complex with a negative imaginary
part. This implies that the real and imaginary parts of $\gamma_0$ are
positive and negative, respectively, and  $\exp[ i (\beta_0 y +
\gamma_0 z)] $ is an amplifying outgoing plane wave  as $z \to +
\infty$. We are concerned with resonant states with only one radiating
plane wave for either $z > D$ or $z < -D$. Therefore, it is assumed
that $\mbox{Im}(\gamma_m) > 0$ for all $m \ne 0$. 
Resonant states form bands where each band corresponds to $k$ being a
complex-valued function of $\alpha$ and $\beta$.
The $Q$ factor of a resonant mode is given by $Q =
-0.5\RE(k)/\IM(k)$.  A BIC is a special
point (with a real $k$) in a band of resonant states.  

We assume the periodic structure has an up-down mirror symmetry, i.e.,
$\varepsilon({\bf r}) = \varepsilon(y,-z)$ for all ${\bf r}$,  then 
it is sufficient to study the field of any eigenmode in the upper half 
space ($z>0$), because the field components are either even in $z$ or
odd in $z$. For a BIC with a frequency and wavevector satisfying Eq.~(\ref{onerc}),  
it is possible to define a topological charge based on the far field
polarization vector of the surrounding resonant states
\cite{zhen14,maksim}.  Let
$\mathcal{C}$ be a closed contour
in the $\alpha$-$\beta$ plane. Each point on
$\mathcal{C}$ corresponds to a resonant 
state in a band that contains the BIC. The resonant state contains a
far-field outgoing plane wave  (for $z \to +\infty$) with a  vector
amplitude ${\bf c}_0^+$. Its projection on the $x$-$y$ plane is 
\begin{equation}
  \label{projected}
  \widehat{\bfm{E}} =
  \insMat{ c^+_{0x} \\ c^+_{0y} \\  0} 
  e^{ i (\alpha x + \beta y + \gamma_0 z)}, 
\end{equation}
where $c^+_{0x}$ and $c^+_{0y}$ are the $x$ and $y$ components of 
${\bf c}_0^+$. For any fixed $z> D$, the real projected electric field
$\mbox{Re} ( \widehat{\bf E} e^{-i \omega t})$ is typically elliptically 
polarized,
and the major polarization vector (along the major axis of the
polarization ellipse) forms an angle  
$\theta$ with the $x$ axis. If it is possible to define $\theta$
continuously as $(\alpha, \beta)$ traverses along $C$ in
counterclockwise direction from a starting
point back to the same point (the ending point), and $\theta|_{\rm end} - \theta|_{\rm
  start} = 2\pi q$ for some $q$,  where $\theta|_{\rm start}$ and $\theta|_{\rm end}$
are the values of $\theta$ at the starting point and the ending point
respectively, then $q$ is the winding number (of the projected major polarization 
vector) on $\mathcal{C}$. Alternatively, $q$ can be evaluated by 
the integral formula
\begin{equation}
  \label{Eq9}
	q = \frac{1}{2\pi}\oint_\mathcal{C} d {\bm \alpha} \cdot
        \nabla_{\bm  \alpha} \theta({\bm \alpha}), 
\end{equation}
where ${\bm \alpha} = (\alpha,\beta)$ and  $\nabla_{\bm \alpha} =
(\partial_\alpha, \partial_\beta)$. 
A BIC is a polarization singularity, since it does not have a far
field (${\bf c}_0^\pm  = 0$). The
topological charge of a BIC is defined as the winding number $q$, if
$\mathcal{C}$ is sufficiently close to the BIC and encloses the BIC in
the $\alpha$-$\beta$ plane  \cite{zhen14,maksim}. 

It is important to note that the major polarization vector or the
angle $\theta$ is undefined, if the projected far-field plane wave is circularly
polarized. This implies that  a circularly polarized state (CPS),
i.e., a resonant state with a  circularly polarized far field, is also
a polarization singularity. If $\mathcal{C}$ contains the wavevector
of a CPS, the winding number on $\mathcal{C}$ is undefined. The
definition of topological charge requires the underlying
assumption that no CPSs exist in a small neighborhood (in the
$\alpha$-$\beta$ plane) of the BIC. 

Finally, we recall that the polarization state of the plane wave given
by Eq.~(\ref{projected}) can be characterized by the Stokes 
parameters~\cite{Stokes}
\begin{align}
\mathcal{S}_0 & = |c_{0x}^+|^2+ |c^+_{0y}|^2,\quad 
  \mathcal{S}_1 = |c_{0y}^+|^2- |c^+_{0x}|^2,\\
\mathcal{S}_2  &= 2 |c^+_{0x} c^+_{0y}| \cos(\varphi_y-\varphi_x),\\
  \mathcal{S}_3 & = 2 |c^+_{0x} c^+_{0y}| \sin(\varphi_y-\varphi_x),
\end{align}
where $\varphi_x = \arg(c^+_{0x} )$, $\varphi_y = \arg(c^+_{0y} )$ are
the phases of $c_{0x}^+$ and $c_{0y}^+$, respectively. Since a BIC has
no radiation, it corresponds to the point $\cbrc{\mathcal{S}_0,\mathcal{S}_1,\mathcal{S}_2,\mathcal{S}_3=0}$. A
CPS satisfies the condition $(\mathcal{S}_1,
\mathcal{S}_2 , \mathcal{S}_3/\mathcal{S}_0)= (0,0, \pm 1)$, where the 
ellipticity $ \mathcal{S}_3/\mathcal{S}_0 = +1$ for the left CPS 
and $-1$ for right CPS. A linearly polarized state is any point with 
$\mathcal{S}_3=0$. The principle value of angle $\theta$
is given by $\theta =   \arg (\mathcal{S}_1 +i \mathcal{S}_2)/2$.

\section{Circular cylinders}
\label{S2a}

\begin{figure}[htb]
  \centering  
  \includegraphics[width=1\linewidth]{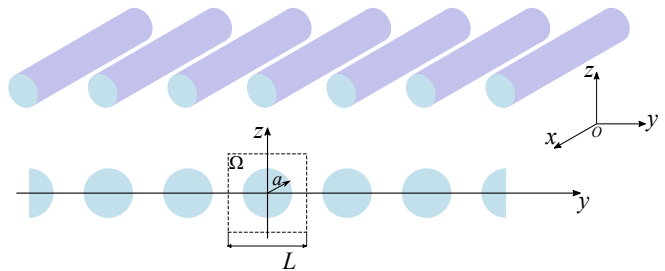}
  \caption{A periodic array of circular dielectric cylinders
    surrounded by air. The array is periodic in $y$ with period
    $L$. The cylinders are parallel to the $x$ axis. The radius 
    and dielectric constant of the cylinders are $a$ and
    $\varepsilon_c$, respectively.}
  \label{fig_carray}
\end{figure}

A periodic array of circular dielectric cylinders surrounded by air,
as shown in Fig.~\ref{fig_carray},
is a simple structure supporting many different BICs~\cite{shipman03,shipman07,bulg14b,yuan17,maksim}.  
In general, a BIC may be a standing wave ($\alpha=\beta=0$), may
propagate along the $y$ axis ($\alpha=0$ and $\beta \ne 0$),
along the $x$ axis ($\alpha \ne 0$ and $\beta = 0$),  or in both $x$
and $y$ directions ($\alpha \ne 0$ and $\beta \ne 0$). Those BICs with
$\alpha=0$ are scalar ones with either the $H$ or $E$
polarization. The BICs with $\alpha\ne 0$ are vectorial ones with
nonzero $E_x$ and $\Ht_x$.
In Table~\ref{Tab1},
 \begingroup 
 \squeezetable 
\begin{table}[htb]	
	\caption{A few BICs in periodic arrays  of circular cylinders
          with radius $a$ and dielectric constant $\vare_c = 15$.} 
	\centering 
	\begin{ruledtabular}
	\begin{tabular}{llcccc  }
		& $a/L$   & $\alpha L$      & $\beta L$ &$k L$& $q$\\\hline 
		BIC$_1$ & $0.45042$ & $0$       & $0$         &$3.512894$&$-1$\\
		BIC$_2$ & $ 0.45042$& $0.584168$& $0$         &$3.523918$&$+1$\\
		BIC$_3$ & $ 0.45$   & $0$       & $0.479401$  &$3.471973$&$+1$\\
		BIC$_4$ &$ 0.3 $    & $1.190401$& $0.96714$   &$3.196556$&$+1$\\
	\end{tabular}
\end{ruledtabular}
	\label{Tab1} 
\end{table}	
\endgroup
we list four BICs for periodic arrays with a fixed dielectric constant $\vare_c = 15$. The topological charges of
the BICs are listed in the last column. More BICs in this periodic
array can be found in Ref.~\cite{maksim}. 

Normally, the resonant states and the BICs are computed by solving an
eigenvalue problem for the Maxwell's equations. To take advantage of
the special geometry of the circular cylinders, we use a semi-analytic
method based on cylindrical wave expansions. Since the structure is
invariant in $x$ and periodic in $y$, it is sufficient to solve the
eigenmodes in one period of the structure, i.e., a 2D domain
$\Omega_{\infty}$ given by $|y| < L/2$ and 
$|z| < \infty$.  Inside $\Omega_\infty$, there is a square  $\Omega$
given by $|y| < L/2$ and $|z| < L/2$. We assume one cylinder is
located at the center of $\Omega$. For given $\alpha$ and $\omega$,
the electromagnetic field in $\Omega$ can be expanded in vectorial cylindrical
waves with unknown coefficients~\cite{yumao}. If $\beta$ is also specified, we can
expand the electromagnetic field in plane waves (also with unknown
coefficients) for $|z| > L/2$. Relating the field at $y=\pm L/2$
by the quasi-periodic condition, assuming even or odd symmetry in $y$,
and imposing continuity conditions at $z= L/2$, we can obtain an
operator $\mathscr{A}$, such that 
\begin{equation}
  \label{EqOP}
  \mathscr{A}(k, \alpha,\beta) \, \mathbf{u}|_{z=L/2} = {\bf 0},
\end{equation}
where ${\bf u}$ is a column  vector for $E_x$ and
$\widetilde{H}_x$, and ${\bf u}|_{z=L/2}$ denotes ${\bf u}$ at $z=L/2$
for $|y| < L/2$. Notice that $\mathscr{A}$ is an operator that  
acts on a vector of two single-variable functions.
Since $\mathscr{A}$ depends on $k$, Eq.~(\ref{EqOP}) is a nonlinear
eigenvalue problem. In practice, $y \in (-L/2, L/2)$ is sampled by $N$ points,
${\bf u}|_{z=L/2}$ becomes a vector of length $2N$, and $\mathscr{A}$ is
approximated by a $(2N) \times (2N)$ matrix. The method can be
extended to the case where the boundary of the cylinders are slightly
and smoothly deformed. Details are given in Appendix. 

For computing resonant states, $\alpha$ and $\beta$ are given, we look
for a complex $k$ such that Eq.~(\ref{EqOP}) has a nontrivial
solution. One possible approach is to solve $k$ from 
\begin{equation}
  \label{EqLam}
  \lambda_1(\mathscr{A}) = 0,
\end{equation}
where $\lambda_1$ is the eigenvalue of $\mathscr{A}$ with the smallest
magnitude. A BIC is a special resonant state with $\IM(k)= 0$. For
propagating BICs, it is more efficient to treat $\alpha$ and/or $\beta$
also as unknowns. We can solve Eq.~(\ref{EqLam}) for a real $k$, a
real $\alpha$, and/or a real $\beta$.

\section{Slightly noncircular cylinders}
\label{S3}

\begin{figure}[htb]
	\centering 
	\includegraphics[width=1\linewidth]{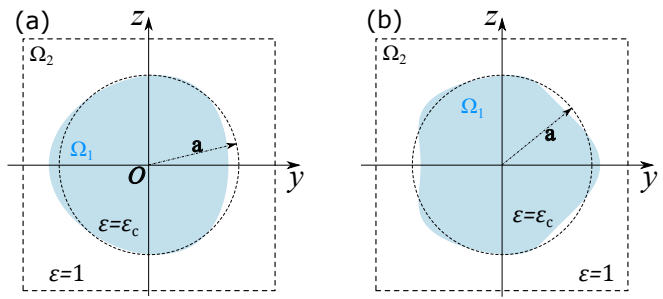}
	\caption{Cross-sections $\Omega_1$ of two deformed cylinders
          with a boundary given by Eq.~\eqref{EqX} for $\delta =
          0.1a$ and the cases (a) $g = 2$ and (b) $g=4$. 
          $\Omega_2$ is the exterior domain outside the cylinder and
          inside the square $\Omega = (-L/2,L/2) \times (-L/2,L/2)$.} 
	\label{Fig2}
      \end{figure}

In this section, we consider a periodic array of  slightly
noncircular cylinders, study the emergence of CPSs when propagating
BICs are destroyed, and also the annihilation and generation of
CPSs. We assume 
the boundary of the cylinder centered at the 
origin is given by
\begin{equation}\label{EqX}
  y = - a\sin(\tau)+\delta \cos(g\tau), \quad 
  z = a\cos(\tau), 
\end{equation} 
for $0\leq \tau < 2\pi$ and $g = 2$ or $4$, where $\delta>0$ is a 
small deformation parameter. The dielectric constant of the cylinders
is fixed at $\vare_c = 15$. The deformed cylinders for $\delta = 0.1a$
are shown in Figs.~\ref{Fig2}(a) and \ref{Fig2}(b)
for $g=2$ and $g=4$, respectively. The small boundary deformation is a
perturbation that breaks the 
reflection symmetry in $y$ (the in-plane inversion
symmetry), but preserves the reflection symmetry in $z$ (the up-down
mirror symmetry).

To calculate vectorial resonant states, we extend the method described
in the previous section. Importantly, a general electromagnetic field
in $\Omega$  
(the square domain containing one cylinder centered at the origin) can
still be expanded in cylindrical waves, although the expansions are
more complicated due to the deformation of the cylinder
boundary. As shown in Appendix, the eigenvalue problem for resonant
states is reduced to Eq.~(\ref{EqOP}), where $\mathscr{A}$ is a $(2N)
\times (2N)$ matrix depending on $k$, $\alpha$ and $\beta$, and $N$
is the number of sampling points for an interval of length $L$. 
It is  highly desirable to compute a CPS without calculating all nearby resonant
states. To find a left or right CPS, we solve a real $\alpha$, a real
$\beta$ and a complex $k$, from Eq.~(\ref{EqLam}) and 
\begin{equation}
  \label{Eq15}
  \mathcal{S}_1 = 0, \quad \mathcal{S}_3/\mathcal{S}_0 = \pm 1, 
\end{equation}
where $\mathcal{S}_0$, $\mathcal{S}_1$ and  $\mathcal{S}_3$ are the
Stokes parameters.



\begin{figure}[htb]
	\centering 
	\includegraphics[width =\linewidth]{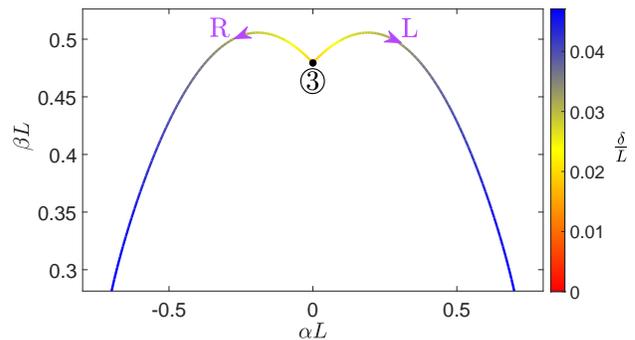}
	\caption{A pair of CPSs with topological charge $1/2$
          emerged from BIC$_3$ for a periodic array of slightly
          noncircular cylinders with $g = 4$.} 
	\label{Fig6}
\end{figure}

\begin{figure}[htb]
	\centering
\includegraphics[width =\linewidth]{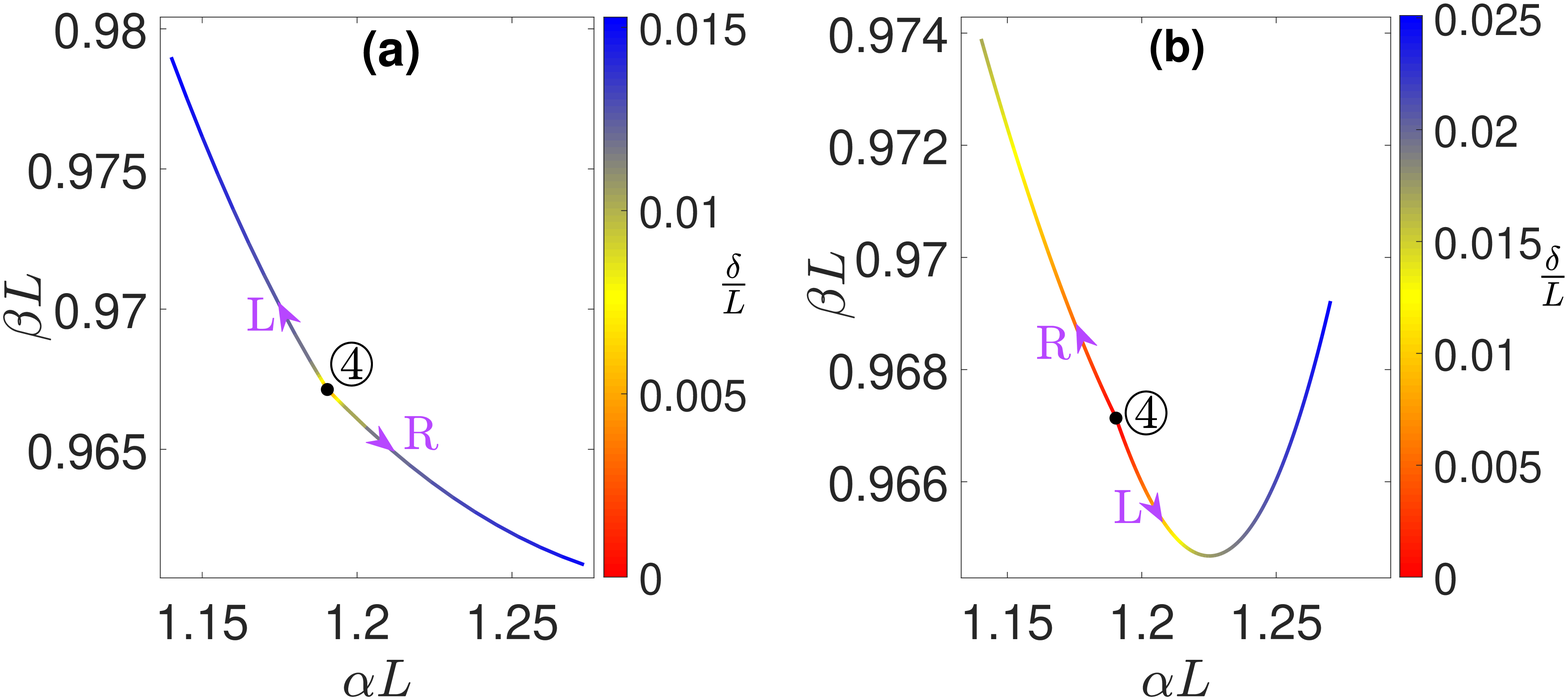}
	\caption{Pairs of CPSs emerged from BIC$_4$ for periodic
          arrays of slightly noncircular cylinders with (a) $g = 2$
          and (b) $g = 4$.} 
	\label{Fig7}
\end{figure}

First, we show that when the deformation parameter $\delta$ is
increased from zero, all four BICs listed in Table~\ref{Tab1} are
destroyed and pairs of CPSs emerge. If the topological charge of the
BIC is 1 (or $-1$), a pair of CPSs with topological charge $1/2$ (or
$-1/2$) and different handedness emerge. The net topological
charge is conserved. 
In Fig.~\ref{Fig6},
we show the emergence of CPSs from BIC$_3$ for a periodic array with 
$a=0.45L$ and $g = 4$. 
The purple arrows indicate the direction of increasing $\delta$. The
curve shows the wavevector $(\alpha, \beta)$ for a pair of left and
right CPSs, and it is shown in color to indicate the value  of
$\delta$. 
These two CPSs exhibit a symmetry with respect to the $\beta$
axis. For each left CPS with wavevector $(\alpha, \beta)$, there is
also a right CPS with wavevector $(-\alpha, \beta)$. The complex
frequencies of these two CPSs are exactly the same. 
In Figs.~\ref{Fig7}(a) and \ref{Fig7}(b), 
we show CPSs emerged from BIC$_4$ for periodic arrays with $g = 2$ and
$g=4$, respectively. The radius of the original cylinders is $a =
0.3L$. 
There is no apparent symmetry between the two CPSs emerged from the
BIC, but there is still a symmetry with respect to the $\beta$ axis. 
Since the structure is invariant in $x$, the resonant states (or BICs) for
$(\alpha, \beta)$ and  
$(-\alpha, \beta)$ are reflections of each other. 
The CPSs with wavevectors $(\alpha, \beta)$ and  $(-\alpha, \beta)$
have the opposite handedness. 

\begin{figure}[htb]
	\centering 
\includegraphics[width =\linewidth]{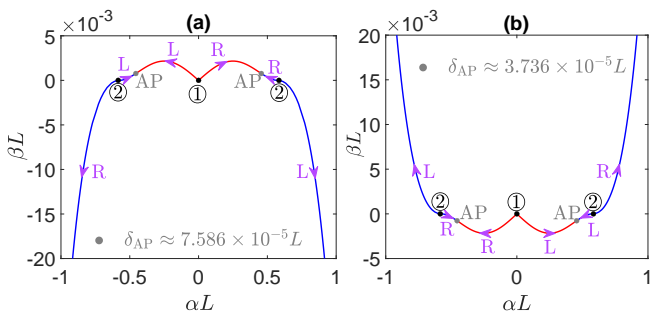}
	\caption{Emergence of CPSs from BIC$_1$ and BIC$_2$, and
          annihilation of CPSs of opposite charges for periodic arrays
          of slightly noncircular cylinders with (a) $g = 2$ and (b) $g = 4$.}
	\label{Fig3}
\end{figure}

Next, we show that as $\delta$ is increased, the CPSs emerged from
BICs of opposite topological charge may collapse to a CPS with a zero
charge. BIC$_1$ and BIC$_2$ in Table~\ref{Tab1} are found in the same
periodic array of circular cylinders with radius $a= 0.45042L$, but
their topological charges are  $-1$ and $+1$, respectively.
As $\delta$ is increased from zero, both BIC$_1$ and BIC$_2$ are 
destroyed. In Figs.~\ref{Fig3}(a) and \ref{Fig3}(b),
we  show the emergence of CPS pairs from BIC$_1$ and BIC$_2$ for
periodic arrays of deformed cylinders with $g =2$ and $g=4$, respectively. 
The CPSs emerged from BIC$_1$ and BIC$_2$ carry
topological charges $-1/2$ and $1/2$, respectively. As $\delta$
reaches a critical value $\delta_{\rm AP}$
($\delta_{\rm AP} \approx  7.586 \times 10^{-5}L$ for $g=2$ and $\delta_{\rm
  AP} \approx  3.736 \times 10^{-5}L$ for $g=4$), two CPSs with the same 
handedness, one from BIC$_1$ and the other from BIC$_2$, collapse to a
CPS with a zero topological charge.  Since these two CPSs cease to
exist for $\delta > \delta_{\rm AP}$, we call $\delta_{\rm AP}$
an annihilation point (AP). 

\begin{figure}[htb]
  \centering 
  \includegraphics[width =\linewidth]{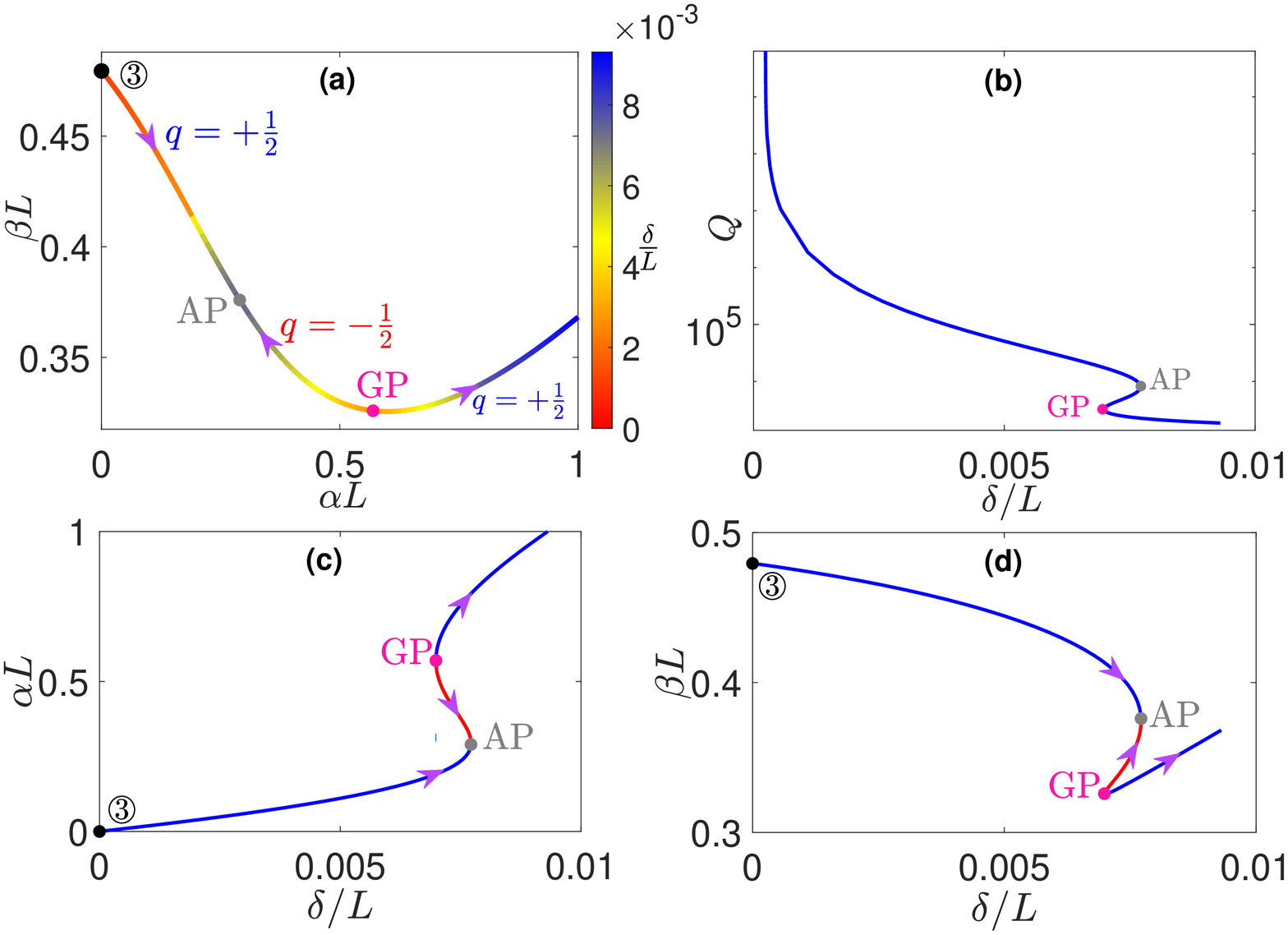}
  \caption{(a) A continuous right CPS emerged from BIC$_3$ for a
    periodic array with $g = 2$, with $\delta$ increased from $0$ to
    $\delta_{AP}$, decreased to $\delta_{GP}$, and increased
    again. The topological charge of the CPS has changed from $1/2$ to
    $-1/2$ and back to $1/2$ accordingly. (b) The $Q$ factor, (c) $\alpha$ and (d) $\beta$ of the right CPS as  multivalued functions of $\delta$. The blue and red curves  
    correspond to topological 
    charge $1/2$ and $-1/2$, respectively. }
  \label{Fig4}
\end{figure}

Finally, we show that CPSs connected to a single BIC may encounter
self-generation and self-annihilation points as $\delta$ is increased.
For a periodic array with $a = 0.45L$ and $g=2$, a continuous branch
of CPSs emerging from BIC$_3$ is shown in Fig.~\ref{Fig4}(a). 
The curve in Fig.~\ref{Fig4}(a) depicts the wavevector
$(\alpha, \beta)$ of a CPS when $\delta$ is first increased from zero
to $\delta_{\rm AP} \approx 0.0077L$, then decreased 
to $\delta_{\rm GP} \approx 0.0070L$, finally increased again. The
topological charge of the CPS has changed from $1/2$ to $-1/2$, and
back to $1/2$ accordingly.
In the process of an increasing $\delta$,  the critical value
$\delta_{\rm AP}$ is annihilation point where two CPSs 
of opposite topological charge and same handedness collapse to a CPS
with a zero charge,  but these two CPSs are not connected
to different BICs. In fact, the CPS with topological charge $-1/2$
and another CPS with topological charge $1/2$ emerge from a CPS of
zero topological charge at $\delta_{\rm GP}$. We call $\delta_{\rm
  AP}$ a self-annihilation point and $\delta_{\rm GP}$ a 
self-generation point. However, annihilation and generation are
relative terms depending on how the structure is tuned. For example,
$\delta_{\rm GP}$ can also be 
regarded as a self-annihilation point, if the structure is tuned by a
decreasing $\delta$.  It is clear that the continuous branch of CPSs
emerged  from BIC$_3$, as shown in Fig.~\ref{Fig4}(a), exhibits a multi-valued
dependence on $\delta$. In
Figs.~\ref{Fig4}(b), (c) and (d),
we respectively show the $Q$ factor, and wavevector components $\alpha$ and $\beta$
as multivalued functions of $\delta$.  Despite the generation and
annihilation of CPSs as $\delta$ is varied, the  net topological
charge  is conserved and remains at $1/2$.  

 \section{Conclusion}
 \label{S4}

Many applications of BICs are realized in periodic structures
sandwiched between two homogeneous media. It is known 
that the existence and robustness of BICs, including the
propagating BICs in periodic structures, depend crucially on
symmetry \cite{hsu13_2,zhen14}. More precisely, it has been proved
that some propagating BICs  
are robust with respect to structural perturbations that preserve the
in-plane inversion symmetry and the up-down reflection symmetry, even
though these BICs do not have a symmetry mismatch with compatible
radiating waves \cite{yuan17ol,conrob}. If the perturbation breaks one of these two
symmetries, the propagating BICs are typically destroyed and become
resonant states with a finite $Q$ factor \cite{hu18}. 
The topological charge of a BIC, defined using the polarization vector
of the resonant states surrounding the BIC in momentum space, is an
interesting concept and is useful for understanding the evolution of BICs as
structural parameters are varied \cite{zhen14,maksim}.
The definition of topological charge does not require the in-plane inversion symmetry. When structural parameters are
varied, the topological charge is always conserved, but this 
does not imply that the BICs are robust with respect to arbitrary
structural perturbations, because a CPS is also a polarization
singularity in momentum space and it can have a nonzero topological
charge. Therefore, the conservation of topological charge is only
valid when both BICs and CPSs are included. For 
propagating BICs, the connection between symmetry-breaking
perturbations and the emergence of CPSs has not been clearly 
established in existing literature. Using a periodic array of slightly
noncircular cylinders as an example, we show that pairs of CPSs emerge when
propagating BICs are destroyed by arbitrarily small perturbations that
break only the in-plane inversion symmetry. We also study the
generation and annihilation of CPSs when structural parameters are
varied. It is shown  that pairs of CPSs of opposite topological charge
can collapse at special CPSs with a zero topological charge, but the net
topological charge is still conserved. 
 
\section*{Acknowledgments}
The authors acknowledge support from the Research Grants Council of
Hong Kong Special Administrative Region, China (Grant No. CityU 11305518).

\appendix*
\section{Construction of matrix $\mathscr{A}$}
        
To obtain a  matrix $\mathscr{A}$ satisfying Eq.~(\ref{EqLam}) for
the case of a periodic array of slightly noncircular cylinders, we
need to find cylindrical wave solutions that are valid inside and outside a 
single cylinder, determine a matrix 
$\mathscr{C}$ mapping ${\bf u}$ to the normal derivative of ${\bf u}$ on
$\partial \Omega$ (the boundary of $\Omega$), and finally construct
the matrix $\mathscr{A}$. 
As shown in Fig.~\ref{Fig2}, $\Omega_1$ is a subdomain of $\Omega$
corresponding to the cross section of the cylinder, $\Omega_2 = \Omega
\backslash \overline{\Omega}_1$,
 $\varepsilon({\bf r}) =\varepsilon_1$ or 
$\varepsilon_2$ for ${\bf r} \in \Omega_1$ or $\Omega_2$,
respectively. The boundary of $\Omega_1$ 
is $\Gamma$. 

For given $k$ and $\alpha$ and an integer $p$, we construct a
vectorial cylindrical wave solution that 
depends on two arbitrary coefficients $c_p$ and $r_p$. The $x$
components of this solution are assumed to be 
\begin{align*}
  E_x^{(p)} &= \begin{cases}
  \displaystyle
  \sum a_{pq}\, \frac{J_q(\rho_1  r)}{J_q(\rho_1  a)}\,e^{iq\theta}, &
  {\bf r} \in\Omega_1\\
\displaystyle	 c_p \frac{J_p(\rho_2  r)}{J_p(\rho_2
                   a)}\,e^{ i p\theta} + \sum b_{pq} \frac{Y_q(\rho_2
                   r)}{Y_q(\rho_2  a)}\,e^{iq\theta},& {\bf r} \in\Omega_2
	\end{cases} \\
\Ht_x^{(p)} &= \begin{cases}
  \displaystyle	\sum  s_{pq}\, \frac{J_q(\rho_1  r)}{J_q(\rho_1
    a)}\,e^{iq\theta}, &   {\bf r} \in\Omega_1\\
\displaystyle	r_p \frac{J_p(\rho_2  r)}{J_p(\rho_2 a)}\,e^{i p\theta} +
        \sum  t_{pq} \frac{Y_q(\rho_2  r)}{Y_q(\rho_2
          a)}\,e^{iq\theta}, & {\bf r} \in\Omega_2
	\end{cases}
\end{align*}
where $r$ and $\theta$ are the polar coordinates of ${\bf r}$, 
$\rho_j = (k^2 \varepsilon_j - \alpha^2)^{1/2}$, $a$ is the
radius of the circle to which $\Gamma$ is close, $J_q$ and $Y_q$ are the first and second kinds of Bessel
functions of order $p$,  the sums are for $q$ from $-\infty$ to
$+\infty$, and $a_{pq}$, $b_{pq}$, $s_{pq}$, $t_{pq}$ (for all $q$) are unknown
  coefficients. 

Let $\nu = (\nu_y,\nu_z)$ be the outward unit vector normal to $\Gamma$, and
$\tau=(-\nu_z,\nu_y)$ be a unit vector tangential to $\Gamma$, then
the tangential field components of this cylindrical wave solution are 
\begin{align}
  \Ht_{\tau}^{(p)} =& \frac{i}{\eta}\brc{\vare \frac{ \partial
                      E^{(p)}_x}{\partial \nu} +\frac{\gamma_0}{k}
                      \frac{ \Ht_x^{(p)}}{\partial \tau}},  \\
E_{\tau}^{(p)} =& \frac{i}{\eta}\brc{ \frac{\partial \Ht_x^{(p)}
                  }{\partial \nu} -\frac{\gamma_0}{k} 
\frac{\partial E_x^{(p)}}{\partial \tau}}.
\end{align}
We discretize $\Gamma$ using $M$ points (assuming $M$ is odd) 
and truncate the sums by $ |q| \leq (M-1)/2$. 
The continuity conditions of $E_x$, $\Ht_x$, $E_\tau$ and
$\Ht_\tau$ at those $M$ points on $\Gamma$ give rise to the following
linear system:
\[
\insMat{ A_{11} & A_{12} & {\bf 0} &  {\bf 0} \cr
{\bf 0}   & {\bf 0}  & A_{23} & A_{24} \cr
  A_{31} & A_{32} & A_{33} & A_{34} \cr
  A_{41} & A_{42} & A_{43} & A_{44} }
\insMat{ {\bf a}_p \cr {\bf b}_p \cr {\bf s}_p \cr {\bf t}_p}
= c_m \insMat{ {\bf g}_1 \cr {\bf 0}  \cr {\bf g}_3 \cr {\bf g}_4}
+ r_m \insMat{ {\bf 0} \cr {\bf h}_2 \cr {\bf h}_3 \cr {\bf h}_4},
\] 
where $A_{jk}$ for $1\le j, k \le 4$ are $M \times M$ matrices, ${\bf
  g}_j$ and ${\bf h}_j$ for $1\le j \le 4$ are column vectors of
length $M$,  ${\bf a}_p$, ${\bf  b}_p$, ${\bf s}_p$ and ${\bf t}_p$
are also column vectors of length $M$, ${\bf a}_p$ is the vector for
all $a_{pq}$,  etc.  
Solving the above linear system, we obtain column vectors
${\bf   f}_{jk}^{(p)}$ for $1 \le j, k \le 2$, such that 
\begin{equation}
{\bf b}_p = c_p {\bf f}_{11}^{(p)} +  r_p {\bf f}_{12}^{(p)}, \quad 
{\bf t}_p = c_p {\bf f}_{21}^{(p)} +   r_p {\bf f}_{22}^{(p)}.
\end{equation}
Therefore, the cylindrical wave solution can be written as
${\bf u}_p = c_p {\bf u}_p^{(1)} + r_p {\bf u}_p^{(2)}$, 
where ${\bf u}_p^{(1)}$ and ${\bf u}_p^{(2)}$ are completely
determined.

The general field in $\Omega$ can be expanded in the above cylindrical
waves as 
\begin{equation}
  {\bf u} = \sum_{p=-\infty}^\infty {\bf u}_p =
  \sum_{p=-\infty}^\infty ( c_p {\bf u}_p^{(1)} + r_p {\bf u}_p^{(2)}).
\end{equation}
If $\partial \Omega$ is sampled by $4N$ points, we can truncate the
index $p$ in the sum above to $-2N  \leq p \leq 2N-1$,  and evaluate ${\bf
  u}$ and the normal  
derivative of ${\bf u}$ at the $4N$ points on $\partial \Omega$. This
leads to $(8N) \times (8N)$ matrices $\mathscr{D}_1$ and
$\mathscr{D}_2$ such that  
\begin{equation}
  {\bf u}|_{\partial \Omega}
  = \mathscr{D}_1 
  \insMat{ {\bf c} \\ {\bf r} },
  \quad
  \left. \frac{\partial {\bf u} }{\partial \nu} \right|_{\partial \Omega}
  = \mathscr{D}_2   \insMat{ {\bf c} \\ {\bf r} }, 
\end{equation}
where ${\bf u}|_{\partial \Omega}$ and $\partial_\nu {\bf
  u}|_{\partial \Omega}$ are column vectors of 
length $8N$, ${\bf c}$ and ${\bf r}$ are column vectors of length $4N$ with 
entries $c_p$ and $r_p$, respectively. For simplicity, $\partial_\nu$
is simply taken to be $\partial_z$ or $\partial_y$ on the horizontal
and vertical sides of $\Omega$. Therefore, we have matrix $\mathscr{C}
= \mathscr{D}_2 \mathscr{D}_1^{-1}$, such that  
\begin{equation}\label{DtN}
  \left. \frac{\partial {\bf u} }{\partial \nu} \right|_{\partial \Omega}
  = \mathscr{C}   {\bf u}|_{\partial \Omega}.  
\end{equation}
If $\beta$ is given, ${\bf u}$ and $\partial_y {\bf u}$  satisfies the
following quasi-periodic condition:
\begin{align}
{\bf u}|_{y=L/2} &= e^{ i \beta L} {\bf u}|_{y=-L/2}, \\
\left. \frac{\partial {\bf u} }{\partial y} \right|_{y=L/2} &= e^{ i \beta L} 
\left. \frac{\partial {\bf u} }{\partial y} \right|_{y=-L/2}.  
\end{align}
Since the structure has the up-down mirror symmetry, we have either 
${\bf u}(y,z) ={\bf u}(y,-z)$ or ${\bf u}(y,z) =-{\bf
  u}(y,-z)$. Substituting these conditions into Eq.~(\ref{DtN}), we
obtain a matrix $\mathscr{B}_0$ such that 
\begin{equation}
  \label{DtN2}
  \left. \frac{\partial {\bf u} }{\partial z} \right|_{z=L/2}
  = \mathscr{B}_0   {\bf u}|_{z=L/2}.  
\end{equation}
For $z > L/2$, the field can be expanded in plane waves as in
Eq.~(\ref{planewave}). Using the plane wave expansion for ${\bf u}$,
evaluating ${\bf u}$ and $\partial_z {\bf u}$ at $z=L/2$, and
eliminating the unknown coefficients, we obtain a matrix
$\mathscr{B}_1$, such that 
\begin{equation}
  \label{DtN3}
  \left. \frac{\partial {\bf u} }{\partial z} \right|_{z=L/2}
  = \mathscr{B}_1  {\bf u}|_{z=L/2}.  
\end{equation}
This leads to Eq.~(\ref{EqOP}) for $\mathscr{A} = \mathscr{B}_1 -
\mathscr{B}_0$.


\begin{thebibliography}{99}


\bibitem{neumann29} J. von Neumann and E. Wigner, 
 ``\"{U}ber   merkw\"{u}rdige diskrete Eigenwerte,'' 
Phys. Z.  {\bf 30},  465-467 (1929).   


\bibitem{hsu16} C. W. Hsu, B. Zhen, A. D. Stone,
  J. D. Joannopoulos, and M. Solja\v{c}i\'{c}, 
 ``Bound states in the continuum,'' 
 Nat. Rev. Mater. {\bf 1}, 16048 (2016).


\bibitem{bonnet94} A.-S. Bonnet-Bendhia and F. Starling, ``Guided
 waves by electromagnetic gratings and nonuniqueness examples for the
 diffraction problem,''  Math. Methods Appl. Sci.  {\bf 17}, 305-338 (1994). 

\bibitem{padd00} P. Paddon and J. F. Young, ``Two-dimensional 
  vector-coupled-mode theory for textured planar waveguides,'' 
\prb\ {\bf 61}, 2090-2101 (2000). 

\bibitem{ochiai01} T. Ochiai and K. Sakoda, ``Dispersion relation and
  optical transmittance of a hexagonal photonic crystal slab,'' 
\prb\ {\bf 63}, 125107 (2001). 

\bibitem{tikh02} S. G. Tikhodeev, A. L. Yablonskii, E. A Muljarov,
  N. A. Gippius, and T. Ishihara, 
``Quasi-guided modes and optical properties of photonic crystal 
slabs,'' \prb\ {\bf 66}, 045102 (2002). 

\bibitem{shipman03} S. P. Shipman and S. Venakides, 
``Resonance and bound states in photonic crystal slabs,'' 
SIAM J. Appl. Math.  {\bf 64}, 322-342 (2003). 

\bibitem{lee12} J. Lee, B. Zhen, S. L. Chua, W. Qiu, J. D. Joannopoulos, 
  M. Solja\v{c}i\'{c}, and O. Shapira, ``Observation and 
  differentiation of unique high-Q optical resonances near zero wave 
  vector in macroscopic photonic crystal slabs,'' \prl\ {\bf 109}, 
  067401 (2012). 

\bibitem{port05} R. Porter and D. Evans, ``Embedded Rayleigh-Bloch 
  surface waves along periodic rectangular arrays,''  Wave Motion 
  {\bf 43}, 29-50 (2005). 

\bibitem{mari08} D. C. Marinica, A. G. Borisov, and 
  S. V. Shabanov, ``Bound states in the continuum in photonics,'' 
  \prl\ {\bf 100}, 183902 (2008).   

\bibitem{hsu13_2} C. W. Hsu, B. Zhen, J. Lee, S.-L. Chua,
  S. G. Johnson, J. D. Joannopoulos, and M. Solja\v{c}i\'{c},
  ``Observation of trapped light within the radiation continuum,'' 
  Nature {\bf 499}, 188--191 (2013). 

\bibitem{bulg14b} E. N. Bulgakov and A. F. Sadreev, ``Bloch 
  bound states in the radiation continuum in a periodic array of 
  dielectric rods,''   \pra\ {\bf 90}, 053801 (2014).

\bibitem{gan16} R. Gansch, S. Kalchmair,  P. Genevet,
  T. Zederbauer, H. Detz,  A. M. Andrews, W. Schrenk, F. Capasso,  M. Lon\v{c}ar, and 
 G. Strasser, ``Measurement of bound states in the continuum by a 
 detector embedded in a photonic crystal,'' Light: Science \&
 Applications {\bf 5}, e16147 (2016). 


  
  \bibitem{evans94}  D. V. Evans, M. Levitin and  D. Vassiliev,
   ``Existence theorems for trapped modes,''
 J. Fluid Mech. {\bf 261}, 21-31 (1994).

\bibitem{bulg08} E. N. Bulgakov and  A. F. Sadreev, 
 ``Bound states in the continuum in photonic waveguides inspired by 
 defects,''    \prb\  {\bf 78}, 075105 (2008). 



 \bibitem{zou15}   C.-L. Zou, J.-M. Cui, F.-W. Sun, X. Xiong, X.-B. Zou, Z.- F. Han, and G.-C. Guo, ``Guiding light through optical bound states in the continuum for ultrahigh-$Q$ microresonantors,''
  Laser Photonics Rev.   {\bf 9}, 114-119  (2015).


\bibitem{bezus18} E. A. Bezus, D. A. Bykov, and L. L. Doskolovich,
  ``Bound states in the continuum and high-$Q$ resonances supported by a dielectric  ridge on a slab waveguide,''
  Photonics Research  {\bf  6}, 1084-1093 (2018).
  
\bibitem{nguyen19} T. G. Nguyen, G. Ren, S. Schoenhardt, M. Knoerzer, A. Boes, and A. Mitchell,
  ``Ridge resonance in silicon photonics harnessing bound states in the continuum,''
  Laser Photonics Rev. {\bf 13},  1900035 (2019).

  \bibitem{yu19} Z. Yu, X. Xi, J. Ma,   H. K. Tsang,  C.-L. Zou, and X. Sun,
  ``Photonic integrated circuits with bound states in the continuum,''
  Optica  {\bf 6}, 1342-1348 (2019).

\bibitem{byk20} D. A. Bykov, E.  A. Bezus, and L. L. Doskolovich,
  ``Bound states in the continuum and strong phase resonances in 
  integrated Gires-Tournois interferometer,'' 
  Nanophotonics {\bf 9}(1), 83-92 (2020).




\bibitem{yuan17_2} L. Yuan and Y. Y. Lu, ``Strong resonances on
  periodic arrays of cylinders and optical bistability with weak
  incident waves,''
  \pra\ {\bf 95}, 023834 (2017).

\bibitem{yuan18} L. Yuan and Y. Y. Lu,
  ``Bound states in the continuum on periodic structures surrounded by
  strong resonances,''
  \pra\ {\bf 97}, 043828 (2018).

  \bibitem{jin19} J. Jin, X. Yin, L. Ni, M. Soljacic, B. Zhen, and C.
  Peng, ``Topologically enabled unltrahigh-$Q$ guided
  resonances robust to out-of-plane scattering,''
  Nature  {\bf 574}, 501-504 (2019).


\bibitem{kosh18} K. Koshelev, S. Lepeshov, M. Liu, A. Bogdanov, and 
  Y. Kivshar, ``Asymmetric metasurfaces with high-$Q$
  resonances governed by bound states in the contonuum,'' 
  \prl\, {\bf      121}, 193903 (2018) 

\bibitem{yuan20} L. Yuan and Y. Y. Lu, ``Perturbation theories for 
    symmetry-protected bound states in the continuum on 
    two-dimensional periodic structures,'' 
    \pra\ {\bf 101}, 043827 (2020). 
  
\bibitem{mocella15} V. Mocella and S. Romano, ``Giant field 
  enhancement in photonic lattices,'' \prb\ {\bf 92}, 155117 
  (2015). 

\bibitem{yoon15} J. W. Yoon, S. H. Song, and R. Magnusson, ``Critical 
  field enhancement of asymptotic optical bound states in the 
  continuum,'' Sci. Rep. {\bf 5}, 18301 (2015).
  
\bibitem{huzhen20} Z. Hu, L. Yuan, and Y. Y. Lu,
  ``Resonant field enhancement near bound states in the continuum on periodic  structures,'' 
  \pra\ {\bf 101}, 043825 (2020).

\bibitem{yuan17} L. Yuan and Y. Y. Lu, ``Propagating Bloch modes 
  above the lightline on a periodic array of cylinders,'' 
  J. Phys. B: Atomic, Mol. and Opt. Phys. {\bf 50}, 05LT01 (2017).


  
  \bibitem{shipman07} S. Shipman and D. Volkov, ``Guided modes in 
  periodic slabs: existence and nonexistence,''  
SIAM J. Appl. Math. {\bf 67}, 687--713 (2007). 

\bibitem{yuan20b} L. Yuan and Y. Y. Lu, ``Parametric dependence of 
    bound states in the continuum on periodic structures,'' 
    \pra\ {\bf 102}, 033513  (2020).

\bibitem{maksim} E. N. Bulgakov and D. N. Maksimov, 
  ``Bound states in the continuum and polarization singularities in 
  periodic arrays of dielectric rods,'' 
  \pra \ {\bf 96}, 063833 (2017). 

\bibitem{hu18} Z. Hu and Y. Y. Lu, ``Resonances and bound states in 
  the continuum on periodic arrays of slightly noncircular 
  cylinders,'' J. Phys. B: At. Mol. Opt. Phys.  {\bf 51}, 035402 
  (2018).   
    
    

\bibitem{zhen14} B. Zhen, C. W. Hsu, L. Lu, A. D. Stone, and M. 
Solja\v{c}i\v{c},  ``Topological nature of optical bound 
states in the continuum,'' 
\prl\ {\bf 113}, 257401 (2014). 

\bibitem{yuan17ol} L. Yuan and Y. Y. Lu,  ``Bound states in the 
  continuum on periodic structures: perturbation theory and 
  robustness,'' \ol\ {\bf 42}(21), 4490-4493 (2017).

\bibitem{conrob} L. Yuan and Y. Y. Lu, ``Conditional robustness of 
  propagating bound states in the continuum on biperiodic 
  structures,'' 
  arXiv preprint arXiv:2001.00832  (2020).



\bibitem{fudan} W. Liu, B. Wang, Y. Zhang, J. Wang, M. Zhao, F. Guan,
  X. Liu, L. Shi, and J. Zi, ``Circularly Polarized States Spawning from
  Bound States in the Continuum,''
  \prl\ {\bf 123}, 116104 (2019).

\bibitem{notomi} T. Yoda and M. Notomi, ``Generation and Annihilation of
  Topologically Protected Bound States in the Continuum and Circularly
  Polarized States by Symmetry Breaking,''
  \prl\ {\bf 125}, 053902 (2020).

  \bibitem{yin20} X. Yin, J. Jin, M. Solja\v{c}i\'{c},  C. Peng, and 
    B. Zhen, ``Observation of topologically enabled unidirectional guided 
    resonances,'' Nature (London)  {\bf 580}, 467-471 (2020). 


\bibitem{Stokes} W. H. McMaster, ``Polarization and the Stokes
parameters,'' Am. J.  Phys. {\bf 22}(6), 351-362 (1954).
    

\bibitem{yumao} Y. Wu and Y. Y. Lu, ``Dirichlet-to-Neumann map method
  for analyzing periodic arrays of cylinders with oblique incident
  waves,'' \josab\   
  {\bf 26}, 1442-1449 (2009). 
  
\end{thebibliography}
\end{document}